\documentclass[10pt ]{revtex4}
\usepackage{hyperref}
\usepackage{amsmath,amsfonts}
\usepackage[dvips]{graphics}
\usepackage[dvips]{graphicx}
\usepackage{subfigure}
\usepackage{epsfig}

\begin{document}
	\title{Stable remnants and quantum gravity effects in non linear electric source Culetu black hole}
	\author{Yawar H. Khan$^{a}$}
	\email{iyhkphy@gmail.com}
	\author{Prince A. Ganai$^{a}$}
	\email{princeganai@nitsri.com}
	\author{Sudhaker Upadhyay$^{b,c,d}$}
	\email{sudhakerupadhyay@gmail.com}

	\affiliation{${}^{a}$Department of Physics, National Institute of Technology, Srinagar, Kashmir-190006, India }
	\affiliation{${}^{b}$Department of Physics, K. L. S. College, Magadh University, Nawada-805110,  India}
	\affiliation{${}^{c}$Visiting Associate, Inter-University Centre for Astronomy and Astrophysics (IUCAA) Pune, Maharashtra-411007}
	 \affiliation{${}^{d}$School of Physics, Damghan University, P. O. Box 3671641167, Damghan, Iran}
	\begin{abstract}
In this paper, we discuss remnants and quantum gravity effects on thermodynamics of new kind of regular black hole known as Culetu black hole. The black hole results in coupling of gravity to non linear electrodynamic source. We first derive various thermodynamic quantities of interest. We then discuss in detail the existence  and stability of remnants. The stability of remnants is analyzed in presence of thermal fluctuations and we observe that thermal fluctuations tend to increase the entropy at the horizon radius which is exactly the remnant radius. We then  investigate non perturbative quantum gravity effects on the thermodynamics of Culetu black hole up to leading order terms.
\end{abstract}
\maketitle
	\section{introduction}
	The general theory of relativity, predicting most celebrated objects known as black holes, has by far now not given an appropriate reason for the existence of these objects. It is by and large accepted that the existence of black holes is due to the presence of singularities in the solutions to Einstein's field equations. These singularities are inevitable under Penrose and Hawking theorems of singularities \cite{i1,i2}. In context, it is also widely believed that these singularities are mere limitations of classical theories of gravity and could be avoided by incorporating the quantum effects. Such a complete description to be known as quantum gravity is yet to be fully established. However, there are some attempts to avoid singularities at a semi-classical level. The first proposed solution of such kind to avoid singularities was given by Bardeen \cite{i3} (known as "Bardeen black hole"). Bardeen in his model replaced the singularity at origin by a de-Sitter patch. Much later, the physical source of Bardeen black hole which is nonlinear electrodynamics of the gravitationally collapsing matter was given by Ayon-Beato and Gracia \cite{i4}. Later on regular black holes with different features like spherical symmetry, axial symmetry, rotation have also been proposed \cite{i5,i6,i7,i8,i9,i10,i11}. In general relativity (GR) one can find singularity-free solutions of the Einstein field equations if the gravity is assumed to be coupled to nonlinear electrodynamics. In \cite{i12} a gravity model coupled to non-linear electric source was derived by Hristu Culetu. The solution in \cite{i12} was obtained by coupling a modified version of the Schwarzschild metric to a nonlinear electric source. The solution thus obtained is a charged regular black hole which is asymptotically Reissner-Nordstrom (RN) type.
	
	It is well known that black holes behave like thermodynamic systems. In reference \cite{i13}, it was shown that the area of the event horizon of a black hole never decreases, which is quite analogous to the positive and ever-increasing value of entropy of any thermodynamical system. This analogy gave rise to the beginning of an important branch of black hole physics termed as black hole thermodynamics. The entropy and the temperature of the black hole are given by its event horizon area $A$ and surface gravity $k$ respectively. Hawking in \cite{i14}, semiclassically proved that black holes can radiate. Bardeen in \cite{i15} proposed four laws of black hole mechanics and this paved way for the interesting field of black hole thermodynamics to flourish. The other thermodynamic quantities like internal energy, free energy, Gibbs energy, specific heat, etc were derived after Hawking and Page \cite{i16} for the first time obtained partition function and ensemble for black holes in AdS (anti-de-sitter) space.
	
	The thermodynamics of black holes can be tanged with an important phenomenon known as thermal fluctuations around equilibrium thermodynamics \cite{i16a,i16b,i16c,i16d}. These thermal fluctuations are stereotypes of the quantum gravity regime and tend to modify the entropy of the black hole. The corrections due to these thermal fluctuations are logarithmic.  The thermal fluctuations to the thermodynamics of rotating AdS black holes and black Saturn have been done in \cite{i16h,i16i,i16j}. With the advent of new models of gravity which incorporated quantum nature of gravity, the thermodynamics of black holes was inevitably changed by a new type of corrections, called quantum correction \cite{i17,i18}. These quantum corrections are non-perturbative and their effect is also logarithmic. The quantum correction (of type $\alpha ln A$) to Godel black hole have been carried out in \cite{i19}. Also in \cite{i20} using Cardy formalism the effect of quantum gravity corrections has been studied. The excellent text of quantum corrections to the thermodynamics of black holes is one given by   Nozari in reference \cite{i19a}. The quantum corrections to the thermodynamics of quasi-topological black holes have been carried by   Upadhyay \cite{i20a}. Some other excellent examples of effect of quantum corrections on thermodynamics of black holes could be found in references \cite{i21,i22,i24,i25,i26,i27,i28,i29,i29a,i29b}. The quantum correction (of type $\alpha ln A$), are generally referred to as quantum one-loop corrections. Another kind of approach that tends to observe the effects of quantum corrections on the entropy of black holes is GUP-corrections to thermodynamics of black holes. The GUP corrections to thermodynamics of black holes have been elegantly discussed in \cite{i29c}. The black hole thermodynamics and correction to the thermodynamics of Horava-Lifitz black hole has also been studied in detail \cite{i30,i31,i32}. A detailed comparison of different approaches of calculating corrections to black hole thermodynamics can be found in reference \cite{i33}. 
	\section{Culetu regular black hole and its non extended phase thermodynamics}
	In this section we briefly introduce the Culetu black hole solution and discuss its thermodynamics in non extended phase space. The action of GR coupled to nonlinear electrodynamics can be written as \cite{cb1,cb1a},
	$$ S =  \int \bigg (\frac{R}{16 \pi}- \frac{1}{4 \pi} L(F)\bigg ) \sqrt{-g}d^4x ,$$ where the Lagrangian $L(F)$  is function of electromagnetic scaler $F$ given by $F= 1/4\pi F^{ab}F_{ab}$. Culetu employed a modified version of Schwarzchild metric to obtain a regular charged black hole solution for the above given action. The metric taken by Culetu is the one given by Xiang et al. \cite{cb2}, which is,
	\begin{equation}
	ds^2= - (1+ 2 \Psi)dt^2+ (1+ 2 \Psi)^{-1} dr^2+r^2d\Omega^2,
	\end{equation} 	where $\Psi(r)= -(m/r)\exp(-\epsilon(r))$ and $d\Omega^2$ refers to unit 2-sphere. The function $\epsilon(r)$ is the factors which helps in removing the singularity. Culetu made a choice of choosing the function $\Psi(r)$ as,
	\begin{equation}
	\Psi(r)= -\frac{m}{r}\exp^{-\frac{k}{r}}.
	\end{equation} With the the quantity $k$ related to charge by the expression $ k= \frac{q^2}{2m}$, our metric function becomes,
	\begin{equation}
	f(r) = 1- \frac{2m}{r} e^{\frac{-q^2}{2mr}}.
	\end{equation} 
	Where $m$ is mass function and is  is related to the mass of black hole. The black hole metric solution we discuss here has asymptotically Minkowski core \cite{cb3}. We now calculate the Hawking temperature $(T_H)$ for Culetu black hole as $T= \frac{1}{4 \pi}f^\prime(r)\bigg|_{r=r_H} $, where $r_H$ denotes larger root of $f(r)=0$ and is the event horizon radius. Thus,
	 \begin{equation} \label{temp}
	 T=  \frac {1} {2\pi r_H}\bigg (1 - \frac {q^2} {2 m r_H}\bigg).
	 \end{equation}
	 The temperature of Culetu black hole solution is positive. This behavior of temperature would be useful in discussing remnants of Culetu black hole, see later sections. We now calculate entropy for Culetu black hole. The entropy can be easily calculated by using Komar energy formula. The entropy obeys Area law and is given by \cite{i12},
	 \begin{equation}\label{enteq}
	 S_0= \pi r_H^2.
	 \end{equation}
	 Here we have denoted entropy by $S_0$, which is equilibrium value of entropy. The essence of this value of entropy would be clear in later sections. In canonical ensemble formalism, using temperature and entropy one can derive almost all thermodynamic quantities of interest. One such thermodynamic quantity of interest is Helmholtz free energy $F$. We calculate $F$ by using classical thermodynamic relation, $F= - \int S_0 dT$, thus we simply have the expression for free energy of Culetu black hole as, 
	 \begin{equation}\label{func}
	 F= \frac{1}{2} \left(r_H-\frac{q^2 \log [r_H]}{m}\right).
	  \end{equation} 
	In the extended phase space the pressure of any black hole system is directly related to cosmological constant. As we have restricted ourselves to non extended phase space, we derive pressure for Culetu black hole thermodynamic system, using classical thermodynamic relation \cite{qc1}, $P= \frac{1}{2}TS$, where $T$ and $S_0$ stand for temperature and entropy of black hole.
	Thus we get the expression for pressure as
	\begin{equation}
	P=\frac{1}{4} r_H \left(1-\frac{q^2}{2 m r_H}\right).
	\end{equation}
	  Another thermodynamic potential of interest is Gibbs free energy. We derive Gibbs free energy using standard relation
	  $G= M- T S_0$, where $M, T, S_0 $ denote mass, temperature and entropy of Culetu black hole. The expression for Gibbs energy is then obtained as
	  \begin{equation} \label{gi}
	  G= \frac{q^2}{2 r_H W\left(\frac{q^2}{r_H^2}\right)}-\frac{1}{2} r_H \left(1-\frac{q^2}{2 m r_H}\right).
	  \end{equation}
   	Here $W$ is the \text{ProductLog} function. We now finally derive specific heat for Culetu black hole thermodynamic system. We would use the relation $C= T\frac{\partial S}{\partial T}$. On substituting for $S_0$ and $T$, we arrive at 
     \begin{equation} \label{sheat}
   C= -\frac{\pi  r_H^2 \left(q^2-2 m r_H\right)}{q^2-m r_H}.
     \end{equation}
     \section{Remnants and their stability in presence of thermal fluctuations}
    The end state of Hawking evaporation results in the formation of stable or meta-stable mass known as black hole remnant. It is widely perceived that when the black hole mass reaches the Plank scale limit, Hawking radiation may stop. This formation of black hole remnant is widely believed to be there as a result of the quantum gravitational nature of the matter near the Planck scale of mass and energy \cite{r0}. The black hole remnants have been widely studied in the literature \cite{r1,r2}. The study of black hole remnants is very important in the context of the information loss paradox \cite{r3}. Also, a stable remnant is believed to be the source of dark energy \cite{r4,r5}. We know from reference \cite{cb3} that the Culetu black hole solution has asymptotically Minkowski core. Also, the temperature for such a class of black holes would be always positive, in contrast to their singular counterparts. This prompts to check the existence of remnants left over at these asymptotically Minkowski cores. With the emission of Hawking radiations, the temperature of the Culetu black hole eventually reaches zero thereby leaving a remnant. We plot temperature for different values of remnant radius as,
    
\begin{figure}[htb]
	\begin{center}$
		\begin{array}{c }
		\includegraphics[width=80 mm]{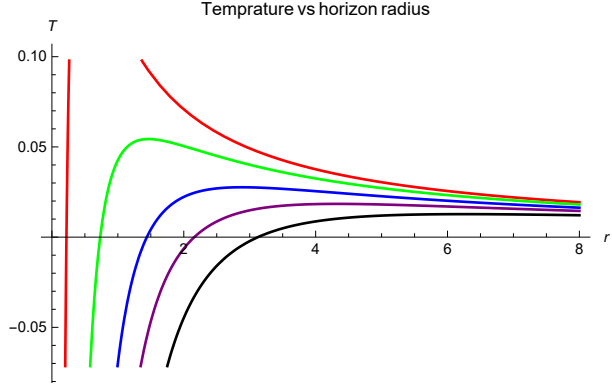}  
		\end{array}$
	\end{center}
	\caption{Temperature vs horizon radius. Here red, green, blue, purple and black lines refer to $ m = .1 \textnormal{ with } q = .21,  m = 1 \textnormal{ with } q = 1.21, m=2\textnormal{ with } q= 2.4, m=3 \textnormal{ with } q=3.6  and m=4 \textnormal{ with } q=5$ respectively.}
	\label{f1}
\end{figure} From the temperature plot, we can see that for each case of specific values $m$ and $q$ the temperature decreases with the increase in horizon radius. For each case, the temperature curve touches the x-axis at different points. These points are radii of remnants for different values of $m$ and $q$. If we decrease the horizon radius further, the temperature becomes negative and these regions are physically forbidden. For higher values of the horizon radius, all the temperature curves show the same behavior and get flattened. We now calculate the radius of black hole remnant by equating equation \ref{temp} to zero. From here we get,
$$ r_{rem}= \frac{q^2}{2 m}.$$ We now analyze the radius of remnant for different values of charge $q$ and mass $m$.
     \begin{center}
    \begin{tabular}{|c|c|c|c|c|c|c|c|c|} 
    		    	\hline 
   	\rule[3ex]{0pt}{4ex} $m$ & .1 & .5 & 1 & 1 & 2 & 2 & 3 & 5 \\ 
   	\hline 
   	\rule[3ex]{0pt}{4ex} $q$ & .1 & .5 & 1 & 1.21 & 2 & 2.4 & 3.6 & 6.05 \\ 
   	\hline 
   	\rule[3ex]{0pt}{4ex} $r_{rem}$ &.025& .25 & .5 & .732&1 & 1.44&2.16 & 3.66    \\ 
   	\hline 
   \end{tabular}
\label{tab}
\end{center}.\\
We can clearly see that size of remnant increases with increase in mass $m$ and charge $q$. The specific case of $m=1$ and $q=1.21$ is at event horizon radius.
\subsection{Thermal corrections to black hole entropy}
\begin{figure}[htb]
	\includegraphics[width=70 mm]{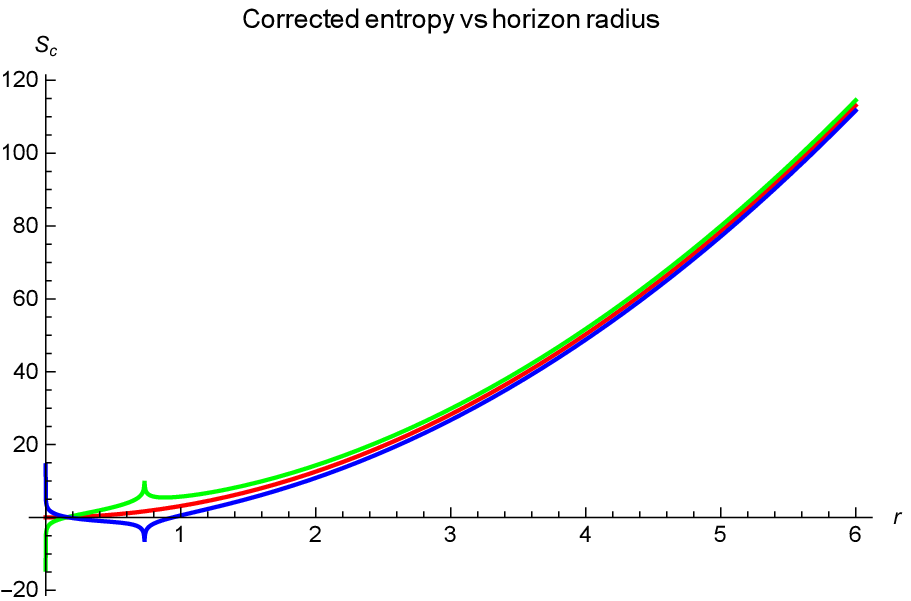}\includegraphics[width=70mm]{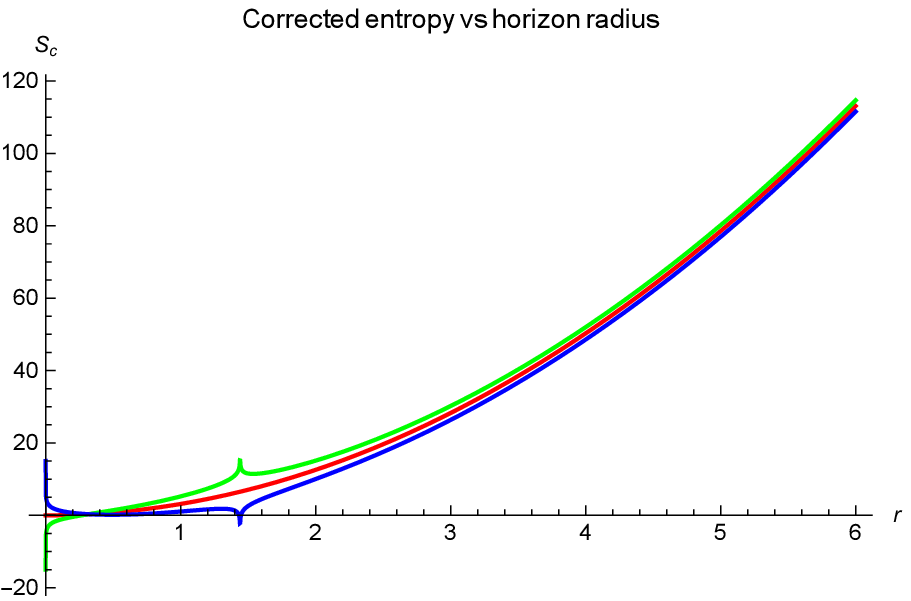} 
	\caption{Corrected entropy versus horizon radius, left plot is for $m$= 1 and $q=1.21$ and the right plot is for $m=2$ and $q= 2.4$. Red, green and blue curves correspond  to $\alpha =0$  and $\alpha=1/2$ and $\alpha$= -1/2 respectively.}
	\label{ecplot}
\end{figure}

	First we briefly discuss leading-order correction to the entropy of Culetu black hole due to thermal fluctuations around equilibrium thermodynamics. Several works in literature suggest that the origin of black hole entropy is statistical in nature \cite{c1,c2,c3,c4,c5}. We employ statistical mechanics of canonical ensemble to derive the corrections to black hole entropy. In canonical ensemble formalism we write the expression for partition function in terms of density of states $\rho (E)$ as follows,
	\begin{equation}
	Z(\beta) = \int_{0}^{\infty} dE \rho (E) e^{-\beta E},
	\end{equation} 
	where $\beta = \frac{1}{T_H}$ with the Boltzmann constant is set to unity. Applying inverse Laplace transformation we estimate density of states as:
	\begin{equation} \label{a}
	\rho(E) = \frac{1}{2\pi i} \int_{\beta_0 - i\infty}^{\beta_0 - i\infty} d\beta Z(\beta)e^{\beta E} = \frac{1}{2\pi i} \int_{\beta_0 - i\infty}^{\beta_0 - i\infty} d\beta e^{S(\beta)},
	\end{equation} 
	where $S(\beta) = \ln Z + \beta E $ is the entropy as function of inverse temperature. We now expand $S(\beta) $ around the equilibrium value of  $\beta$ (i.e. $\beta_0$) via Taylor series expansion. This results,
	\begin{equation} \label{y}
	S(\beta) = S_0 +\frac{1}{2}(\beta - \beta_0)^2 \left. \frac{d^2S}{d\beta^2}\right|_{\beta = \beta_0} + \textnormal{higher-order terms}.
	\end{equation}
	Now substituting the expansion \ref{y} in \ref{a}, we get,
	\begin{equation} \label{denf}
	\rho(E) = \frac{e^{S_0}}{\sqrt{2\pi{\frac{d^2S}{d\beta^2}}}}.
	\end{equation}
	Now taking logarithm in \ref{denf} we get the corrected entropy as,
	\begin{equation} \label{w}
	S= \ln(\rho) = S_0 - \frac{1}{2} \ln \frac{d^2S}{d\beta^2} + \textnormal{other sub-leading terms.}
	\end{equation}
	The second term in the RHS of above equation is identified as correction term. This correction term in a region of positive specific heat can be evaluated as, 
	\begin{equation}
	\frac{d^2S}{d\beta^2} = CT^2.
	\end{equation}
	Also it is easy to see that, in the regions of positive specific heat, the specific heat of any thermodynamic system either coincides or remains proportional to the equilibrium value of canonical entropy. Thus we have,
	\begin{equation} \label{q}
	S= S_0 - \frac{1}{2} \ln S_0T^2.
	\end{equation}
	We  can use $\alpha$ in place of $\frac{1}{2}$  in the leading-order correction term of equation \ref{q}. The constant $\alpha$ is there to estimate and characterize the effect of thermal fluctuations on various thermodynamic quantities. We can now write the corrected entropy of the  thermodynamic system due to thermal fluctuations as, 
	\begin{equation} \label{scorec}
	S= S_0 - \alpha \ln S_0T^2.
	\end{equation}
	We can observe that the nature of leading order correction term due to thermal fluctuations is logarithmic, which is in par with the findings of \cite{kaul,carlip}. \\
	We now calculate the corrected entropy for Culetu black hole by plugging equations \ref{enteq} and \ref{temp} in equation \ref{scorec}, and we get,
	\begin{equation} \label{entcor}
	S= \pi  r_H^2-\alpha  \log \left[\frac{\left(1-\frac{q^2}{2 m r_H}\right)^2}{4 \pi } \right].
	\end{equation}
	To analyze the   nature and effect of the  correction on entropy, we plot entropy vs horizon radius for different values of correction parameter $\alpha$ in figure \ref{ecplot}.
 	We observe from figure \ref{ecplot} that uncorrected entropy which is shown by the red line is a monotonic increasing function of horizon radius. The corrected entropy for the positive value of correction parameter $\alpha$, which is given by the green curve has a peak, which shows that entropy at this point tends to acquire maximum value. This is the region of the most stable configuration. In the left plot, which corresponds to $m=1$ and $q=1.21$ the green curve has a peak at $r=.732$, which is exactly the radius of remnant for these values of $m$ and $q$ (see the table above).  When we change to $m=2$ and $q=2.4$, the peak of the green curve also shifts. Now the maximum entropy point is at $r_H=1.4$, which is exactly the remnant radius ($r_{rem}$) for $m=2$ and $q=2.4$. Also, it is to be noted that the green curve for smaller values of the horizon radius has a downward shift and goes asymptotic to the negative Y-axis when $r_H$ tends to zero. This means that the positive value of correction parameter $\alpha$ which assigns maximum entropy to black hole remnant has the tendency to decrease entropy and hence preventing the formation of stable structures below the remnant radius. This effect would be more clear when we would discuss the effect of thermal fluctuations on various thermodynamic quantities later. In contrast, if we assign a negative value to the correction parameter $\alpha$, we observe that the entropy has a downward peak at the horizon radius equal to the remnant radius. We say that the negative value of the correction parameter $\alpha$ prevents the formation of a black hole remnant. However, at very small values of the horizon radius, we see that the blue curve has an asymptotic shift towards the positive Y-axis. This is fully opposite what we had observed for the positive value of $\alpha$, which is the green curve. This means that although for a negative value of correction parameter thermal fluctuations prevent the formation of black hole remnant, they try to give stable configurations for very small values of horizon radius $r_H$. Now with this increase in entropy due to thermal fluctuations at remnant radius, we expect that all the other thermodynamic quantities would tend to acquire their maximum stable value at the remnant radius. The remnant thus formed would be the most stable configuration of this black hole system. The stability of remnants is a prerequisite for their formation. In the following subsection, we would discuss the effect of thermal fluctuations of various thermodynamic potentials, particularly near the remnant radius.
\subsection{Effect of thermal corrections on various thermodynamic quantities near remnant radius}
\begin{figure}[htb]
	\begin{center}$
		\begin{array}{c }
		\includegraphics[width=80 mm]{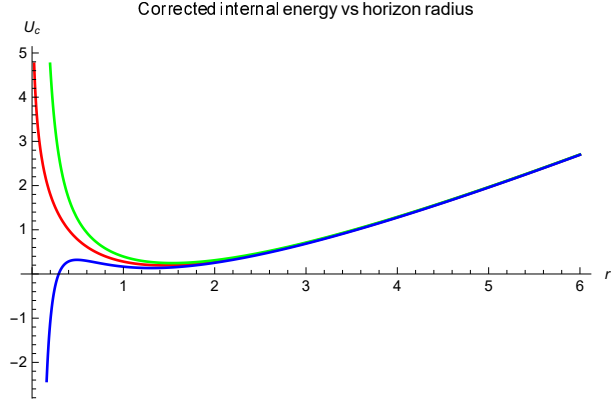}  
		\end{array}$
	\end{center}
	\caption{Internal energy vs horizon radius for $m$=2 and $q=2.4$. The red, green and blue curves are for $\alpha$= 0,.5 and -.5 respectively.}
	\label{ucorpol}
\end{figure}

\begin{figure}[htb]
	\begin{center}$
		\begin{array}{c }
		\includegraphics[width=80 mm]{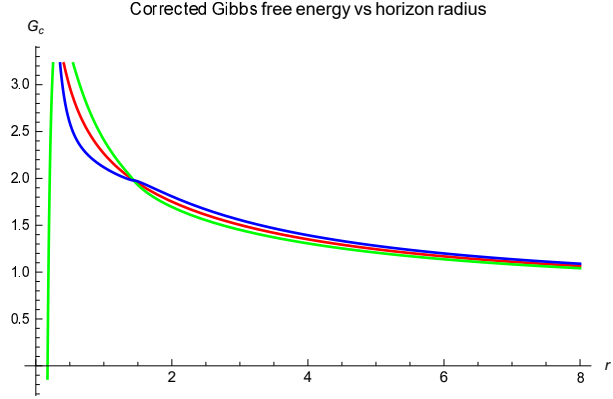}  
		\end{array}$
	\end{center}
	\caption{Corrected Gibbs free energy vs horizon radius for $m$=2 and $q=2.4$. The red, green and blue curves are for $\alpha$= 0, .5 and -.5 respectively.}
	\label{gcplot}
\end{figure}

\begin{figure}[htb]
	\begin{center}$
		\begin{array}{c }
		\includegraphics[width=80 mm]{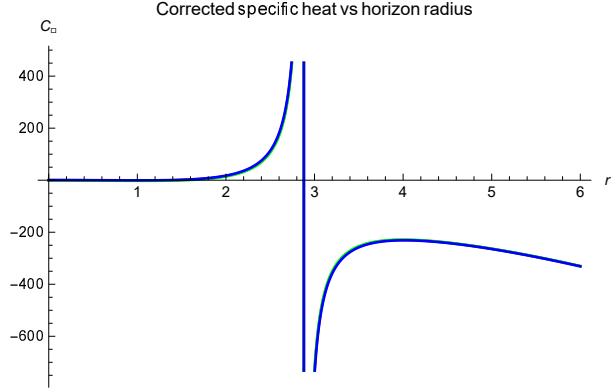}  
		\end{array}$
	\end{center}
	\caption{Corrected specific heat vs horizon radius for $m$=2 and $q=2.4$. The red, green and blue curves are for $\alpha$= 0, .5 and -.5 respectively.}
	\label{shcp}
\end{figure}

	After studying the effects of thermal fluctuations on the entropy of Culetu black hole, we are now in a position to derive all other thermodynamic potentials. We would try to realize the effect of thermal fluctuations on various thermodynamic quantities particularly at the remnant radius. Our aim is to see how in presence of thermal fluctuations various quantities behave near the remnant radius and make remnant the most stable configuration. First, we discuss the Helmholtz free energy or simply free energy. We derive free energy as $F_c= - \int S_0 dT$, where $F_C$ is corrected free energy, $S$ and $dT$ follow from equation \ref{entcor} and \ref{temp} respectively. Thus we have,
	\begin{equation}
	F_c=\frac{-\alpha \left(q^2-2 m r_H\right) \log \left [\frac{\left(1-\frac{q^2}{2 m r_H}\right)^2}{4 \pi }\right ]+2 \pi  m r_H^3+\alpha  q^2-2 \pi  q^2 r_H^2 \log [r_H]}{4 \pi  m r_H^2}.
	\end{equation} For $\alpha \to 0 $, one would get the uncorrected free energy given in equation \ref{func}. We now plot $F_C$ w.r.t horizon radius for different values of correction parameter $\alpha$.
	\begin{figure}[htb]
	\begin{center}$
			\begin{array}{c }
			\includegraphics[width=80 mm]{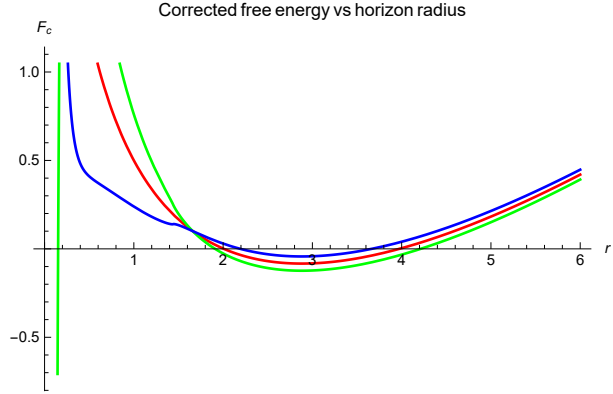}  
			\end{array}$
	\end{center}
		\caption{Free energy vs horizon radius for $m$=2 and $q=2.4$. The red, green and blue curves are for $\alpha$= 0,.5 and -.5 respectively.}
	\label{freeplot}
	\end{figure}
From figure \ref{freeplot} we can see that the positive value of $\alpha$ which increases the entropy at the horizon radius decreases the free energy near the horizon radius. This decrease in free energy is an indication of stability for any thermodynamic system. Thus the positive value of the correction parameter of thermal fluctuations decreases the free energy and hence makes the remnant stable. However, below the remnant radius, the green curve is well above all the other lines. This means that the positive value of the correction parameter does not allow the formation of stable configurations below the remnant radius. This is the same as what we had observed from the entropy plot. In contrast, the negative value of the correction parameter $\alpha$ has a tendency to minimize the free energy below the horizon radius. This is due to the positive asymptotic shift of entropy curve for negative values of correction parameter $\alpha$, see blue line in figure \ref{ecplot}.\\
Next, we discuss the behavior of internal energy in presence of thermal fluctuations near the remnant radius. The expression for corrected internal energy denoted by $U_c$ is given by,
\begin{equation}
U_c= \frac{1}{4} \left[\frac{q^2 \left(\frac{\alpha }{\pi  r_H^2}-1\right)}{m}-\frac{2 q^2 \log [r]_H}{m}+4 r_H\right].
\end{equation} 
The plot of internal energy vs horizon radius for different values of correction parameter is given in figure \ref{ucorpol}.
 From figure \ref{ucorpol}, it is clear that near horizon radius equal to remnant radius, all the three curves try to merge. This indicates that thermal fluctuations do not alter the equilibrium condition of the black hole. This can be easily understood using first law of black hole mechanics \cite{i15}. We know from first law of black hole mechanics that a black hole thermodynamic system could be always thought as equilibrium between black hole mass and thermal radiations (Hawking radiation). Any change in internal energy would mean some radiations either going in or out of the black hole body. This clearly indicates that the remnant we discuss here is not only made stable by thermal fluctuations but is also in equilibrium condition. \\
Another thermodynamic potential of interest is Gibbs free energy. The uncorrected Gibbs free energy of Culetu black hole thermodynamic system is given in equation \ref{gi}. The expression for corrected Gibbs free energy is derived by replacing the entropy $S_0$ by $S$ and we get,
\begin{equation}\label{gce}
G_c= \frac{q^2}{2 r_H W\left(\frac{q^2}{r_H^2}\right)}-\frac{\left(1-\frac{q^2}{2 m r_H}\right) \left(\pi  r_H^2-\alpha  \log \left[\frac{\left(1-\frac{q^2}{2 m r_H}\right)^2}{4 \pi }\right]\right)}{2 \pi  r_H}.
\end{equation}The plot corresponding to equation \ref{gce}, in which $G_c$ is plotted against horizon radius $r_H$ for different values of correction parameter $\alpha$ is given in figure \ref{gcplot}. It is evident from the plot that the positive value of correction parameter $\alpha$, shown by the green line decreases the Gibbs free energy near the remnant radius. This decrease in Gibbs free energy of any thermodynamic system including black holes is evidence of stability. We know from the corrected entropy plots given in figure \ref{ecplot}, that the positive value of correction parameter $\alpha$ hinders the formation of stables structures below the remnant radius. This is also clear from Gibbs free energy plot, where we see that below the remnant radius the green line is above the red line. This means that thermal fluctuations for the positive value of correction parameter $\alpha$ try to take the system into unstable regions by increasing  Gibbs free energy. In contrast, below the remnant radius, the blue line which is for the negative value of correction parameter $\alpha$ is at a minimum among all other curves. This tendency of negative values of correction parameter $\alpha$, of inducing stability for regions below remnant radius is evident from entropy plot \ref{ecplot}. It is necessary to emphasize that the uncorrected Gibbs free energy of Culetu black hole for parameters $m=2$ and $q=2.4$ is positive. This means that this system has a natural tendency to undergo a phase transition from unstable to stable configurations. 
We now discuss the effect of thermal fluctuations on the specific heat of Culetu black hole near the remnant radius. The specific heat of the Culetu black hole in absence of thermal fluctuations is given in expression \ref{sheat}. Upon incorporating the corrected entropy formula in the derivation of \ref{sheat}, we arrive at the  expression for corrected specific heat in presence of thermal fluctuation as,
\begin{equation}
C_c= \frac{\pi  r_H^2 \left(q^2-2 m r_H\right)+\alpha  q^2}{m r_H-q^2}.
\end{equation}
We now plot the corrected specific heat with respect to horizon radius for different values of correction parameter $\alpha$ in figure \ref{shcp}.
From figure \ref{shcp} it is evident that thermal fluctuations do not change the heat content of the Culetu black hole thermodynamic system. All three lines red, green, and blue which correspond to $\alpha$ = 0, .5, and -.5 have merged into a single line. Also, a discontinuity at $r_H$= 2.75 (approx) means that the black hole system undergoes a phase transition from stable configuration to unstable, which is small to larger (SBH-LBH) black hole transition. This also means that remnants that we have discussed for $m=2$ and $q=2.4$ are stable remnants by virtue of positive specific heat. Any remnants with a radius greater than $r_H=2.75$ would be unstable. A single discontinuity in all the specific heat curves indicates that the phase transition occurring would be of the first order. 
\section{Quantum corrections to thermodynamics of Culetu black hole}
\begin{figure}[htb]
	\begin{center}$
		\begin{array}{c }
		\includegraphics[width=80 mm]{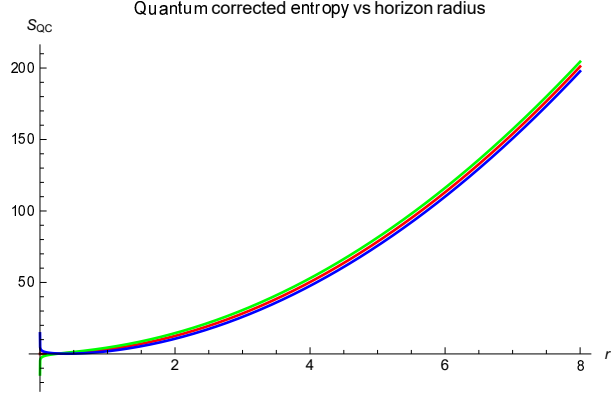}  
		\end{array}$
	\end{center}
	\caption{Quantum corrected entropy vs horizon radius for $m$=2 and $q=2.4$. The red, green and blue curves are for $\mu$= 0, .5 and -.5 respectively.}
	\label{qcplo}
\end{figure}

\begin{figure}[htb]
	\begin{center}$
		\begin{array}{c }
		\includegraphics[width=80 mm]{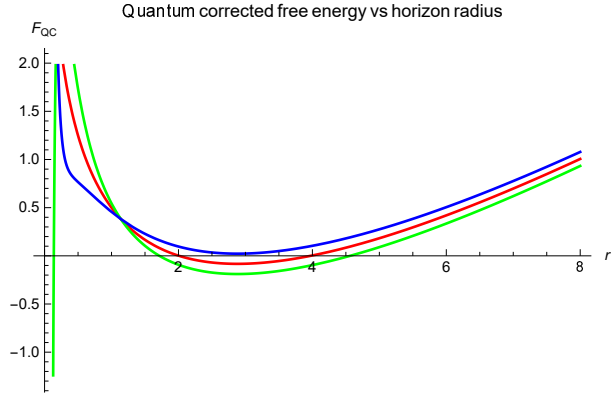}  
		\end{array}$
	\end{center}
	\caption{Quantum corrected free energy vs horizon radius for $m$=2 and $q=2.4$. The red, green and blue curves are for $\mu$= 0, .5 and -.5 respectively.}
	\label{qcf}
\end{figure}

\begin{figure}[htb]
	\begin{center}$
		\begin{array}{c }
		\includegraphics[width=80 mm]{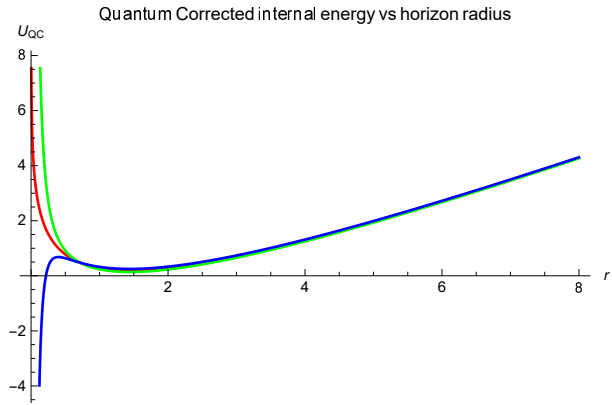}  
		\end{array}$
	\end{center}
	\caption{Quantum corrected Internal energy vs horizon radius for $m$=2 and $q=2.4$. The red, green and blue curves are for $\mu$= 0, .5 and -.5 respectively.}
	\label{Uqc}
	\end{figure}

\begin{figure}[htb]
	\begin{center}$
		\begin{array}{c }
		\includegraphics[width=80 mm]{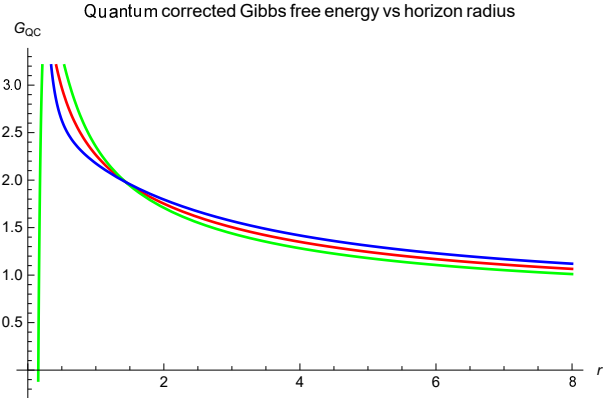}  
		\end{array}$
	\end{center}
	\caption{Quantum corrected Gibbs free energy energy vs horizon radius for $m$=2 and $q=2.4$. The red, green and blue curves are for $\mu$= 0, .5 and -.5 respectively.}
	\label{gibplot}
\end{figure}

Apart from thermal fluctuations, a non-perturbative correction in quantum gravity regime can also change the entropy of any black hole system. The leading order terms in  these kind of correction are also logarithmic. Using approach of \cite{qc1}  and references there in, we find the expression for corrected entropy of Culetu black hole as,
\begin{equation}\label{enqc}
S_{QC}= S_0 + \mu \log A.
\end{equation}Here $S_0$ is equilibrium value of entropy, $\mu$ is the correction parameter and $A$ is area of event horizon. While importing equation \ref{enqc}, we have changed the symbol $\alpha$ to $\mu$. Also we mention that we have restricted ourselves to only first order corrections. The value of the correction parameter $\mu$ depends upon the approach we adopt to study the corrections. Thus quantum corrected entropy in our case would be,
\begin{equation}\label{sqc}
S_{QC}= \pi r_H^2+ \mu \log [4 \pi r_H^2].
\end{equation}We plot the corrected entropy given in equation \ref{sqc} w.r.t horizon radius for different values of correction parameter $\mu$ in figure \ref{qcplo}. It is evident from figure \ref{qcplo}, that the quantum corrections tend to change the entropy at very small values of the horizon radius. These changes due to quantum corrections happen at Plank scales of length and energy.
 The positive value of the correction parameter $\mu$ tends to make entropy negative, which is physically meaningless. However, the negative value of the correction parameter $\mu$ tends to increase the entropy at smaller values of the horizon radius. This behavior of entropy curves for different values of correction parameter $\mu$ would be clear when we will analyze other thermodynamic quantities. Using the quantum corrected entropy formula and temperature, we derive the expression for quantum corrected free energy of Culetu black hole as,
\begin{equation}
F_{QC}= \frac{\mu  \log[ 4 \pi  r_H^2 ] \left(q^2-2 m r_H\right)+2 \pi  m r_H^3-4 \mu  m r_H+\mu  q^2-2 \pi  q^2 r_H^2 \log [r_H]}{4 \pi  m r_H^2}.
\end{equation}The plot for quantum corrected free energy is drawn in figure \ref{qcf}.
 From the plot of quantum corrected free energy we observe that the negative value of correction parameter $\mu$, which tends to increase entropy for very small values of horizon radius now decreases free energy for smaller values of horizon radius $r_H$. This lowering of free energy for any thermodynamic system is an indication of stability. For sufficiently higher values of horizon radius $r_H$, we see all the three curves red, green, and blue merge. This indicates that quantum corrections do not affect the thermodynamics of large-sized black holes. We would now analyze the effect of quantum corrections on the internal energy of Culetu black hole. Upon incorporating the corrected thermodynamic variables, we derive the expression of quantum corrected internal energy as,
\begin{equation}
U_{QC}= -\frac{\left(\pi  r_H^2-\mu \right) \left(q^2-4 m r_H\right)+2 \pi  q^2 r_H^2 \log [r_H]}{4 \pi  m r_H^2}.
\end{equation}In figure \ref{Uqc}, we plot the quantum corrected internal energy w.r.t horizon radius for different values of correction parameter $\mu$. The plot in figure \ref{Uqc} shows that for the negative value of correction parameter $\mu$ has tendency to decrease the internal energy for small values of horizon radius. This decrease in internal energy is evidence of stability. One can easily understand this from first law of black hole mechanics \cite{i15}, where the black
holes system could always thought to be in equilibrium with black body radiations only at zero temperature. Now if the internal energy is decreasing, we say that the temperature is going towards zero. This indicates that the our black hole system is trying to attain equilibrium between black hole mass and black body radiations (Hawking radiations).
\\
Another thermodynamic quantity of interest is the Gibbs free energy. The expression for quantum corrected Gibbs free energy is as, 
\begin{equation}
G_{QC}=\frac{q^2}{2 r_H W\left(\frac{q^2}{r_H^2}\right)}-\frac{\left(1-\frac{q^2}{2 m r_H}\right) \left(\mu  \log \left[4 \pi  r_H^2\right]+\pi  r_H^2\right)}{2 \pi  r_H}.
\end{equation}
The plot of quantum corrected Gibbs free energy w.r.t horizon radius for different values of correction parameter $\mu$ is given in figure \ref{gibplot}. We can see from the plot of Gibbs free energy that the negative value of correction parameter $\mu$ tends to decrease the Gibbs free energy for smaller values of horizon radius $r_H$ (see blue curve). However, for large values of the horizon radius, all the curves show the same behavior. The quantum corrections which arise at small length scales (Plank scale), always are predominant for smaller values of $r_H$. Thus for large values of $r_H$, our curves show the same behavior. 
\section{Conclusion}
In this paper, we studied the existence and stability of the remnants of Culetu black hole. The black hole has a non-linear electric source and is regular everywhere. Firstly we briefly introduced the non-linear electric source Culetu black hole and studied its non-extended phase thermodynamics. In section III we studied the remnants of Culetu black hole in detail. The stability of these remnants was then analyzed in presence of thermal fluctuations. These fluctuations are small statistical fluctuations around equilibrium thermodynamics. We observed that due to thermal fluctuations the entropy of the Culetu black hole shows a peak towards positive axis at a point which is exactly the remnant radius (for particular values of $q$ and $m$). For different values of $q$ and $m$, the remnant size is different and the peak in the entropy curve due to thermal fluctuations is also correspondingly at different values of horizon radius, see plots in figure \ref{ecplot}. This means that thermal fluctuations try to assign maximum value to the entropy of remnant and make it stable by virtue of the generalized second law of black hole thermodynamics. Once knowing that the entropy of the remnant is increased by thermal fluctuations, we argue that all other thermodynamic quantities should tend to acquire maximum stable values at the remnant radius. Expectedly we found that all thermodynamic quantities like free energy, internal energy, Gibbs free energy, and specific heat deposited there most stable value at a horizon radius which is the same as the remnant radius. The stability of the remnants so discussed is also seen from the plot of specific heat in figure \ref{shcp}. From the plot, we see that specific heat remains positive near the remnant radius. Also, we see from the specific heat plot that for $m = 2$ and $q=2.4$ any remnant of radius $r \leq 2.75$ would have positive specific heat and would be stable. But our calculations show that by virtue of thermal fluctuations the only stable remnant possible for these particular values of $m$ and $q$ is the remnant of radius $r= 1.44$, see table \ref{tab} and figure \ref{ecplot}. In section IV we discussed the effect of quantum corrections on various thermodynamic quantities of Culetu black hole. These quantum corrections are non-perturbative and have a logarithmic effect on thermodynamics of Culetu black hole. We see that the plots of different quantum corrected thermodynamic quantities and those of thermal fluctuations are nearly the same. This is because both corrections enter to the entropy equation as logarithmic terms.

\end{document}